\documentclass{PoS}
\usepackage{amsmath,amssymb}
\usepackage{graphicx}

\title{Interquark potential for the charmonium system with almost physical quark masses}

\ShortTitle{Interquark potential for the charmonium system with almost physical quark masses}

\author{\speaker{Taichi Kawanai}\\%
       Department of Physics, The University of Tokyo, Tokyo 113-0033, Japan\\
       RIKEN-BNL Research Center, Brookhaven National Laboratory Upton, NY 11973-5000, USA\\
       Department of Physics, Brookhaven National Laboratory Upton, NY 11973-5000, USA\\
       E-mail: \email{kawanai@nt.phys.s.u-tokyo.ac.jp}}

\author{{Shoichi Sasaki}\\%
       Department of Physics, The University of Tokyo, Tokyo 113-0033, Japan\\
       E-mail: \email{ssasaki@phys.s.u-tokyo.ac.jp}}

\abstract{
We study an interquark $\QQbar$ potential for the charmonium system, that is determined from
the the equal-time and Coulomb gauge $\QQbar$ Bethe-Salpeter (BS) wavefunction through the effective Schr\"odinger equation.
 This novel approach enables us to evaluate a kinetic heavy quark mass $m_Q$ and 
 a proper interquark potential at finite quark mass $m_Q$, which receives all orders of $1/m_Q$
 corrections on the static $\QQbar$ potential from Wilson loops, simultaneously. 
 Precise information of the interquark potential for both charmonium and bottomonium states 
 directly from lattice QCD provides us a chance to improve quark potential models, where
 the spin-independent interquark potential is phenomenologically described by the Cornell potential
 and the spin-dependent parts are deduced within the framework of perturbative QCD, 
 from first-principles calculations.
 In this study, calculations are carried out in both quenched and dynamical fermion simulations.
 We first demonstrate that the interquark potential at finite quark mass 
 calculated by the BS amplitude method smoothly approaches the 
 conventional static heavy quark potential from Wilson loops 
 in the infinitely heavy quark limit within quenched lattice QCD simulations.
 Secondly, we determine both spin-independent and -dependent parts of the interquark potential 
 for the charmonium system in 2+1 flavor dynamical lattice QCD using the PACS-CS gauge configurations
 at the lightest pion mass, $M_{\pi}=156$ MeV.  
}

\FullConference{ The XXIX International Symposium on Lattice Field Theory - Lattice 2011\\
July 10-16, 2011\\
Squaw Valley, Lake Tahoe, California}

\newcommand\QQbar{Q\overline{Q}}
\newcommand\mQ{m_{Q}}
\newcommand\SdotS{{\mathbf S}_Q\cdot{\mathbf S}_{\overline{Q}}}

\newcommand\SQ{{\mathbf S}_Q}
\newcommand\SQbar{{\mathbf S}_{\overline{Q}}}
\newcommand\tsrc{t_{\rm s}}

\begin{document}

\section{Introduction}
The linearly rising interquark potential in QCD
plays an essential role in the formation of hadrons.
Indeed, the quarkonium states such as charmonia and bottomonia
can be described well by quark 
potential models~\cite{Eichten:1975ag,Godfrey:1985xj,Barnes:2005pb}, where
the Coulomb plus linear potential, the so-called Cornell potential, is phenomenologically adopted
as the spin-independent central potential.
One of the major successes of lattice QCD is to demonstrate that
the static Wilson loop gives a confining potential between infinitely heavy 
quark~($Q$) and antiquark~($\overline{Q}$), which support
the phenomenology of confining quark interactions in the heavy $Q\overline{Q}$ system~\cite{Bali:2000gf}.

As for the spin-dependent potential, 
spin-spin, tensor and spin-orbit terms of the interquark potential can be identified
as relativistic corrections to the static $Q\overline{Q}$ potential, which are
classified in powers of $1/\mQ$ within a framework called potential
non-relativistic QCD (pNRQCD)~\cite{Brambilla:2004jw}.
However, although the spin-spin potential have been precisely calculated 
as one of the next-leading order corrections in quenched lattice QCD~\cite{Koma:2006fw}, 
their attractive spin-spin potential does not even qualitatively agree with 
the corresponding one in quark potential models, where
the repulsive spin-spin interaction is phenomenologically required by heavy quarkonium spectroscopy.
The spin-spin interaction calculated within the Wilson loop formalism
seems to yield wrong mass ordering among hyperfine multiplets~\cite{Koma:2006fw}.
In potential models, functional forms of the spin-dependent terms
are basically determined by perturbative one-gluon exchange~\cite{Godfrey:1985xj}.
Such the Fermi-Breit type potential, which appears at the order of $1/\mQ^2$ within pNRQCD,
would have validity only at short distances and also in the vicinity of $\mQ = \infty$.
The $1/\mQ$ expansion is not formally applicable at the charm quark mass.
%Thus, there is no theoretical input of the proper 
%spin-dependent potential {\it at finite quark mass}.
Therefore, properties of higher-mass charmonium states predicated
in potential models may suffer from large uncertainties in this sense.

Under these circumstances, we have succeeded to determine a proper interquark potential
{\it at finite quark mass} from the equal-time and Coulomb gauge Bethe-Salpeter (BS)
amplitude through an effective Schr\"odinger equation~\cite{{Kawanai:2011xb},{Kawanai:2011jt}}.
(See also Ref.~\cite{Ikeda:2011bs}).
In this proceedings, we will discuss the quark mass dependence of the
spin-independent interquark potential in quenched lattice QCD~\cite{Kawanai:2011xb}
and  then show results of both spin-independent and -dependent parts of the charmonium potential 
in 2+1 flavor full QCD with almost physical quark masses~\cite{Kawanai:2011jt}.
\section{Formulation}
Let us briefly review the new method utilized here to
calculate the interquark potential with the  finite quark mass.
As the first step, we consider the following equal-time $\QQbar$
BS wavefunction in the Coulomb gauge for the
quarkonium states. %~\cite{{Velikson:1984qw},{Gupta:1993vp}}:
\begin{equation}
  \phi_\Gamma({\bf r})= \sum_{{\bf x}}\langle 0| \overline{Q}
  ({\bf x})\Gamma Q({\bf x}+{\bf r})|
  \QQbar;J^{PC}\rangle \label{eq_phi},
\end{equation}
where ${\bf r}$ is the relative coordinate of two quarks 
and $\Gamma$ is any of the 16 Dirac $\gamma$ matrices.
Practically, the BS wavefunction can be extracted from
the following four-point correlation function
\begin{eqnarray}
  && \sum_{{\bf x},{\bf x}^{\prime}, {\bf y}^{\prime}}
  \langle 0  |\overline{Q}({\bf x}, t)\Gamma Q({\bf x}+{\bf r}, t)
  \left(\overline{Q}({\bf x}^{\prime},\tsrc)
  \Gamma Q({\bf y}^{\prime},\tsrc) \right)^{\dagger}
  |0\rangle\nonumber \\
  &&=   \sum_{{\bf x}}\sum_{n} A_n\langle 0|\overline{Q}({\bf x})\Gamma Q({\bf x}+{\bf r})
  |n \rangle e^{-M^\Gamma_n(t-\tsrc)}\label{eq_correlator} \ \ \ %\nonumber\\
   \xrightarrow{t\gg \tsrc} \ \ \ A_0 \phi_\Gamma({\bf r})
  e^{-M^\Gamma_0(t-\tsrc)},
\end{eqnarray}
at the large Euclidean time from source location ($\tsrc$).
Here both quark and anti-quark fields at $\tsrc$ are
separately averaged in space as wall sources.
$M^\Gamma_n$ denotes a mass of the $n$-th quarkonium state $|n\rangle$ in
a given $J^{PC}$ channel.  
For instance, $\Gamma$ is chosen to be $\gamma_5$ and $\gamma_i$
to obtain the rest mass of the pseudo-scalar (PS) state $(J^{PC}=0^{-+})$
and vector (V) state $(J^{PC}=0^{--})$, respectively. %,

The BS wavefunction satisfies the Schr\"odingier
equation with a non-local potential $U$~\cite{Ishii:2006ec}
\begin{equation}
  -\frac{\nabla^2}{\mQ}\phi_\Gamma({\bf r})+
  \int d{\bf r}'U({\bf r},{\bf r}')\phi_\Gamma({\bf r}')
  =E_\Gamma\phi_\Gamma({\bf r}),
  \label{Eq_schr}
\end{equation}
where $\mQ$ denotes the quark kinetic mass.
The energy eigenvalue $E_\Gamma$ of
the stationary Schr\"odinger equation is supposed to be $M_\Gamma-2\mQ$.
If the relative quark velocity $v=|{\nabla}/\mQ|$ is small as $v \ll 1$,
the non-local potential $U$ can generally expand
in terms of the velocity $v$ as
$U({\bf r}',{\bf r})=                                                                               
 \{V(r)+V_{\text{S}}(r)\SdotS+V_{\text{T}}(r)S_{12}+                                                
 V_{\text{LS}}(r){\bf L}\cdot{\bf S} + \mathcal{O}(v^2)\}\delta({\bf r}'-{\bf r})$,
 where $S_{12}=(\SQ\cdot\hat{r})(\SQbar\cdot\hat{r})-\SdotS/3$
 with $\hat{r}={\bf r}/r$, ${\bf S}=\SQ+\SQbar$
 and ${\bf L} = {\bf r}\times (-i\nabla)$~\cite{Ishii:2006ec}.
 Here, $V$, $V_{\text{S}}$, $V_{\text{T}}$ and $V_{\text{LS}}$ represent
 the spin-independent central, spin-spin, tensor and spin-orbit potentials,
 respectively.

In this study, we focus only on the $S$-wave states. 
%($0^{-+}$ and $1^{--}$ states).  
Thus we perform an appropriate projection
 with respect to the discrete rotation, which provides
 the BS wavefunction projected in the $A^{+}_{1}$ representation. 
 The Schr\"odinger equation for the projected
 BS wavefunction $\phi_{\Gamma}(r)$ is reduced to
 \begin{equation}
   \left\{
   - \frac{\nabla^2}{\mQ}
   +V(r)+\SdotS V_{\text{S}}(r)
   \right\}\phi_{\Gamma}(r)=E_\Gamma \phi_{\Gamma}(r)
   \label{Eq_pot}
 \end{equation}
 at the leading order of the $v$-expansion. 
 The spin operator $\SdotS$ can be easily replaced by expectation values.
 As a result, both spin-independent and -dependent
 interquark potentials can be separately evaluated
 through a linear combination of Eq.(\ref{Eq_pot}) calculated
 for PS and V channels as
 \begin{eqnarray}
   V(r)
   &=& E_{\text{ave}}+\frac{1}{\mQ}\left\{
   \frac{3}{4}\frac{\nabla^2\phi_\text{V}(r)}{\phi_\text{V}(r)}+
    \frac{1}{4}\frac{\nabla^2\phi_\text{PS}(r)}{\phi_\text{PS}(r)}
   \right\} \label{Eq_potC}\\
   V_{\text{S}}(r)
   &=& E_{\text{hyp}} + \frac{1}{\mQ}\left\{
  \frac{\nabla^2\phi_\text{V}(r)}{\phi_\text{V}(r)}
  - \frac{\nabla^2\phi_\text{PS}(r)}{\phi_\text{PS}(r)} \right\},\label{Eq_potS}
 \end{eqnarray}
  where $E_{\text{ave}}=M_{\text{ave}}-2\mQ$
 and $E_{\text{hyp}}=M_\text{V}-M_\text{PS}$.
 The mass $M_{\text{ave}}$ denotes the spin-averaged mass as
 $\frac{1}{4}M_\text{PS}+\frac{3}{4}M_\text{V}$.
The derivative $\nabla^2$ is defined by the discrete Laplacian
 with nearest-neighbor points.
 
 Note here that quark kinetic mass $\mQ$ is essentially required 
 in the definition of the interquark potentials.
 Under a simple, but reasonable assumption as 
 $\lim_{r \to \infty} V_S(r) = 0$, 
 which implies that there is no long-range correlation 
 and no irrelevant constant term in the spin-dependent potential,
 we can obtain the quark kinetic mass from the following formula,
\begin{equation}
\mQ
   =  \lim_{r \to \infty}\frac{-1}{E_{\text{hyp}}}\left\{
  \frac{\nabla^2\phi_\text{V}(r)}{\phi_\text{V}(r)}
  - \frac{\nabla^2\phi_\text{PS}(r)}{\phi_\text{PS}(r)} \right\}.
   \label{Eq_quarkmass}
 \end{equation}
%
%
%As a result, one can self-consistently determine both the spin-independent and 
%spin-dependent $\QQbar$ potentials, and also the quark kinetic mass 
%within a single set of four-point correlation functions.
%
%
\section{Numerical results}
We have performed lattice QCD simulations in both quenched and full QCD in this study.
%The results will be described in the following
%subsections, separately.

\subsection{$N_f = 0 $ quenched QCD simulation}
 \begin{figure}
   \centering
   \includegraphics[width=.48\textwidth]{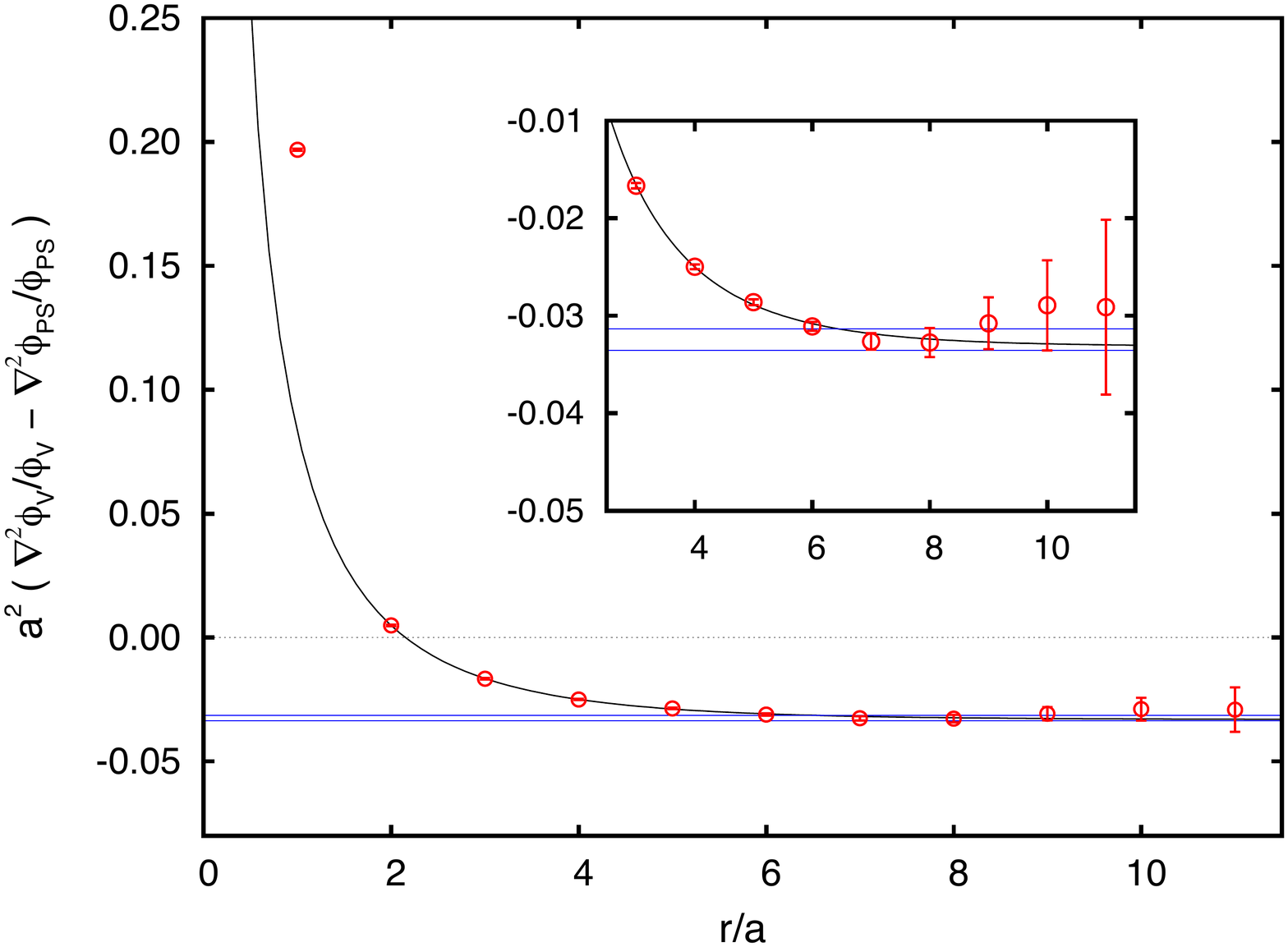}
   \includegraphics[width=.48\textwidth]{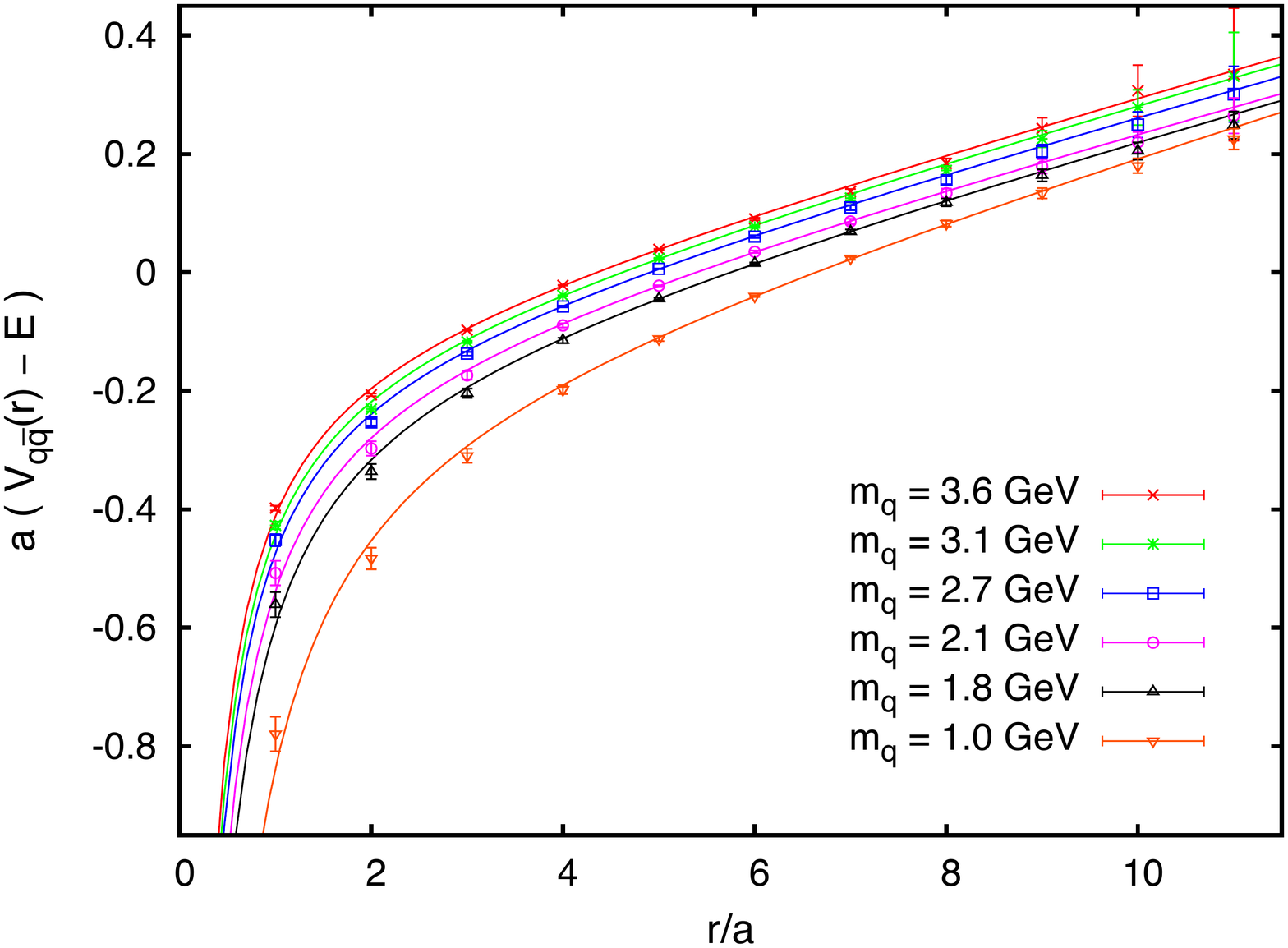}
   \caption{(left) A typical result of
   $\nabla^2\phi_{\rm V}/\phi_{\rm V}-\nabla^2\phi_{\rm PS}/\phi_{\rm PS}$
   as a function of spacial distance $r$. %,
   (right) The interquark potential calculated from the $\QQbar$ BS amplitude
   at finite quark masses covering the range from 1.0 to 3.6~GeV.
   Each curve represents the fit result with the Cornell parametrization.
  }
   \label{Pot_quench}
   \end{figure}
\begin{figure}
\centering
\includegraphics[width=.45\textwidth]{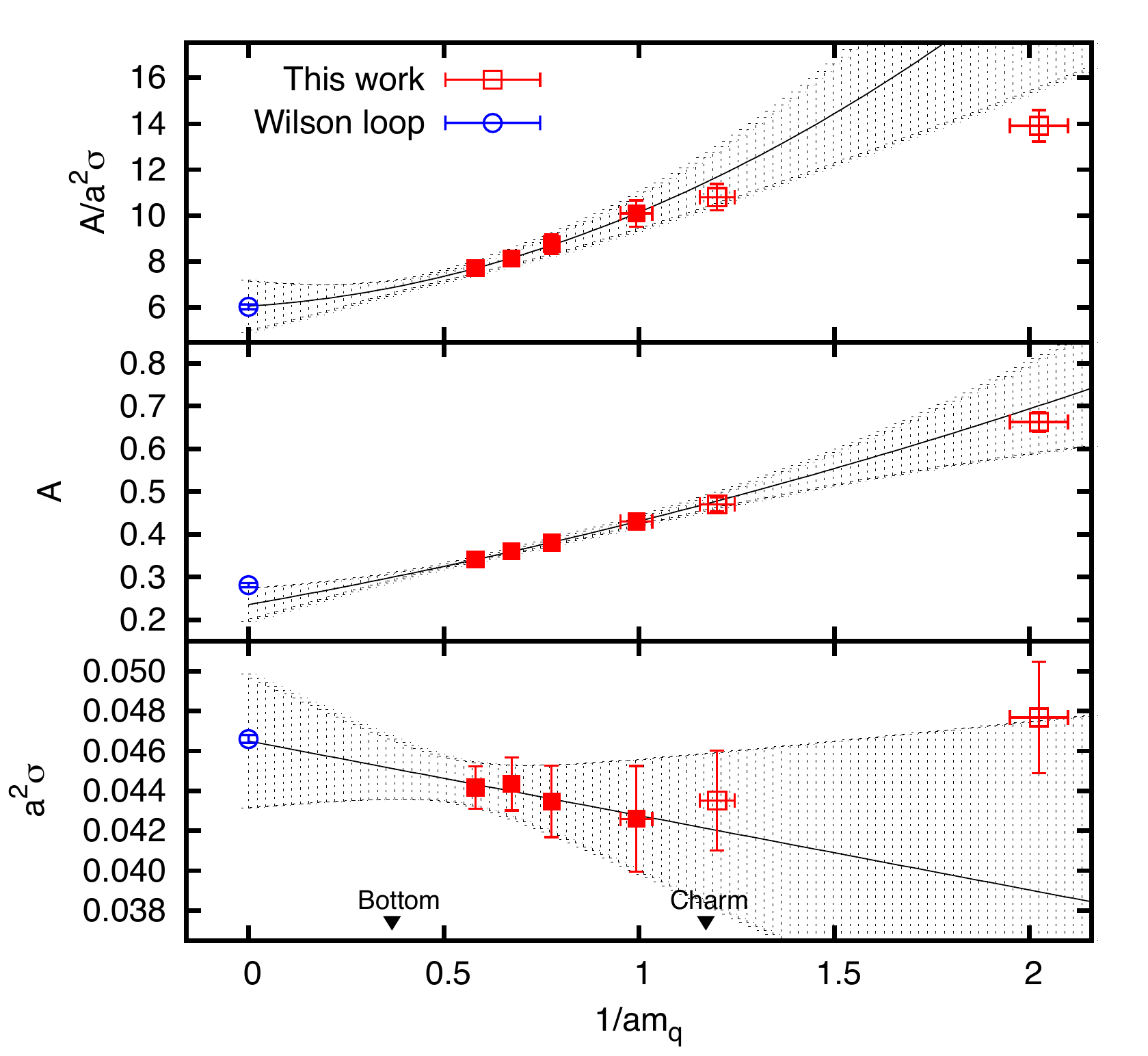}
\caption{
The quark-mass dependence of $A/\sigma$, $A$, and $\sigma$
as functions of $1/\mQ$. We perform the extrapolation towards
the $m_Q \rightarrow \infty$ limit (solid curves) of $A/\sigma$, $A$,  and $\sigma$
with a simple polynomial function in $1/\mQ$.
For $\sigma$, a linear fit with respect to $1/\mQ$ is enough
to describe the data with reasonable $\chi^2/{\rm d.o.f}$, while
quadratic fits are used for $A/\sigma$ and $A$.
The results given by Wilson loops
are also included as open circles. .
\label{para}}
\end{figure}
In quenched lattice QCD simulations, we use a lattice size of $L^3 \times T = 32^3 \times 48$ 
with the single plaquette gauge action at $\beta = 6/g^2 = 6.0$, which corresponds 
to a lattice cutoff of $a^{-1} \approx 2.1$ GeV.
The spatial lattice size corresponds to $La \approx 3\;{\rm fm}$.
 We fix the lattice to Coulomb gauge.
 The heavy-quark propagators are computed using the relativistic heavy-quark (RHQ) action
 with relevant one-loop coefficients of the RHQ~\cite{{Aoki:2001ra},{Kayaba:2006cg}}, 
which  can remove large discretization errors introduced by large quark mass.
  To examine the infinitely heavy-quark limit,
 we adopt the six values of the hopping parameter $\kappa$,
 which cover the range of the spin-averaged mass of $1S$ quarkonium states
 $M_{\rm ave}=\frac{1}{4} (M_{\rm PS} + 3 M_{\rm V})=1.97$-5.86 GeV.
We calculate quark propagators with a wall source located at $\tsrc/a=4$.
  Our results are analyzed on 150 configurations for every hopping parameters.
  
  First, in Fig.~\ref{Pot_quench}, we plot a difference of ratios of
  $\nabla^2\phi_{\rm V}/\phi_{\rm V}$ and $\nabla^2\phi_{\rm PS}/\phi_{\rm PS}$
  as a function of spatial distance $r$ at $\kappa=0.10190$, which is close to the charm quark mass,
  as a typical example. The ratios of $\nabla^2\phi_{\Gamma}/\phi_{\Gamma}$ are
  evaluated by a weighted average of data points in the
  range of $(t-t_{\rm src})/a=$~21~-~23.
  At a glance, the value of $\nabla^2\phi_{\rm V}/\phi_{\rm V}-\nabla^2\phi_{\rm PS}/\phi_{\rm PS}$\
  certainly reaches a nonzero constant value at large distances, which turns out to be
  the value of $-m_Q\Delta E_{\rm hyp}$.
  We then obtain the quark kinetic masses from
  the long-distance asymptotic values of
  $\nabla^2\phi_{\rm V}/\phi_{\rm V}-\nabla^2\phi_{\rm PS}/\phi_{\rm PS}$
  divided by the measured hyperfine splitting $\Delta E_{\rm hyp}$.
  
  Using the quark kinetic mass determined here, we can properly calculate
  the spin-independent interquark potential  from the BS wavefunctions.
  Figure~\ref{Pot_quench} displays results of our potential
   obtained at several quark masses. For clarity of the figure,
  the constant energy shift $E_\text{ave}$ is not subtracted. 
  The resulting $\QQbar$ potentials at finite quark masses exhibit
  the linearly rising potential at large distances and the Coulomb-like potential at short distances
  as originally reported in Ref~\cite{Ikeda:2011bs}.
  
  We simply adopt the Cornell parametrization for fitting our data 
  $V(r)=-A/r+\sigma r + V_0$                                      
  with the Coulombic coefficient~$A$, the string tension $\sigma$,
  and a constant $V_0$.
  All fits are performed over the range $1\le r/a \le 11$.
  In Fig.~\ref{para}, we show the quark-mass dependence of $A/\sigma$, $A$ and $\sigma$
  as functions of $1/\mQ$. First, regardless of the definition of $\mQ$,
  the ratio of $A/\sigma$ in the top figure indicates that the $\QQbar$ 
  potential calculated from the BS wavefunction smoothly approaches the potential 
obtained from Wilson loops in the infinitely heavy-quark limit.
  If we pay attention to the quark-mass dependence of each of the Cornell parameters separately,
  we observe that,  although the Coulombic parameter $A$ depends on
  the quark mass significantly, there is no appreciable dependence of the quark mass
  on the string tension $\sigma$. 
  Their extrapolated values at $m_Q \rightarrow \infty$ are again consistent with those of the
  Wilson loop result.
  The extrapolation curve are also displayed as a solid curve in Fig.~\ref{para}.
\subsection{$N_f = 2+1$ full QCD simulation} 
\begin{table*}%[H]                                   
\centering                                               
  \caption{Summary of RHQ parameters calibrated for charmonium system in dynamical QCD simulation.
      \label{Tab2}
      }
      \begin{tabular}{cccccccc} \hline  \hline                                                     
       &$\kappa_c$ & $\nu$ & $r_s$  & $c_B$ & $c_E$ & \\ \hline
       RHQ parameters  & 0.10819 & 1.2153 & 1.2131  & 2.0268 & 1.7911 & \\  \hline \hline
      \end{tabular}
\end{table*}
  Full QCD simulations are also carried out by using 2+1 flavor gauge
 configurations generated by PACS-CS Collaboration on lattice of size $32^3\times 64$
 with the Iwasaki gauge action at $\beta=1.9$, which corresponds to a comparable
 lattice cutoff of $a^{-1} \approx 2.2$ GeV ~\cite{Aoki:2008sm}.
 The spatial lattice size corresponds to $L \approx 3\;{\rm fm}$.
 The hopping parameters for the light sea quarks
 \{$\kappa_{ud}$,$\kappa_{s}$\}=\{0.13781, 0.13640\}
 correspond to $M_\pi=156(7)$ MeV and $M_K= 554(2)$ MeV~\cite{Aoki:2008sm}.
 Our results are analyzed on all 198 gauge configurations. 
 We also employ RHQ action to compute the heavy quark propagator,
 which has five parameters $\kappa_c$, $\nu$, $r_s$, $c_B$ and $c_E$.
 The parameters $r_s$, $c_B$ and $c_E$ are determined
 by tadpole improved one-loop perturbation theory~\cite{Kayaba:2006cg}.
 For $\nu$, we use a non-perturbatively determined value,
which is adjusted as $c^2_{\text{eff}}=1$ in the dispersion relation
 $E^2({\bf p}^2)=M^2+c^2_{\text{eff}}|{\bf p}|^2$
 for the spin-averaged $1S$ charmonium state. %~\cite{Namekawa:2011wt}.
 We choose $\kappa_c$ to reproduce the experimental spin-averaged mass
 of $1S$ charmonium states $M_\text{ave}^{\text{exp}}=3.0678(3)$ GeV. 
 As a result, the relevant speed of light, $c^2_{\text{eff}}=1.04(5)$, 
 and the  spin-averaged $1S$ charmonium mass, $M_\text{ave}=3.0702(9)$ GeV, 
 is observed with our RHQ parameters summarized in Table~\ref{Tab2}.
 To increase statistics for BS wavefunction,
 we have computed charm quark propagators with two wall sources located
 at different time slices $\tsrc/a=6$ and 57
 and fold them together to create a single four-point correlation function.
 Then the measured hyperfine mass splitting $M_\text{hyp}=0.1137(8)$ GeV
 is close to  the experimental value $M_\text{hyp}^{\text{exp}}=0.1166(12)$ GeV.

 \begin{figure}
   \centering
   \includegraphics[width=.48\textwidth]{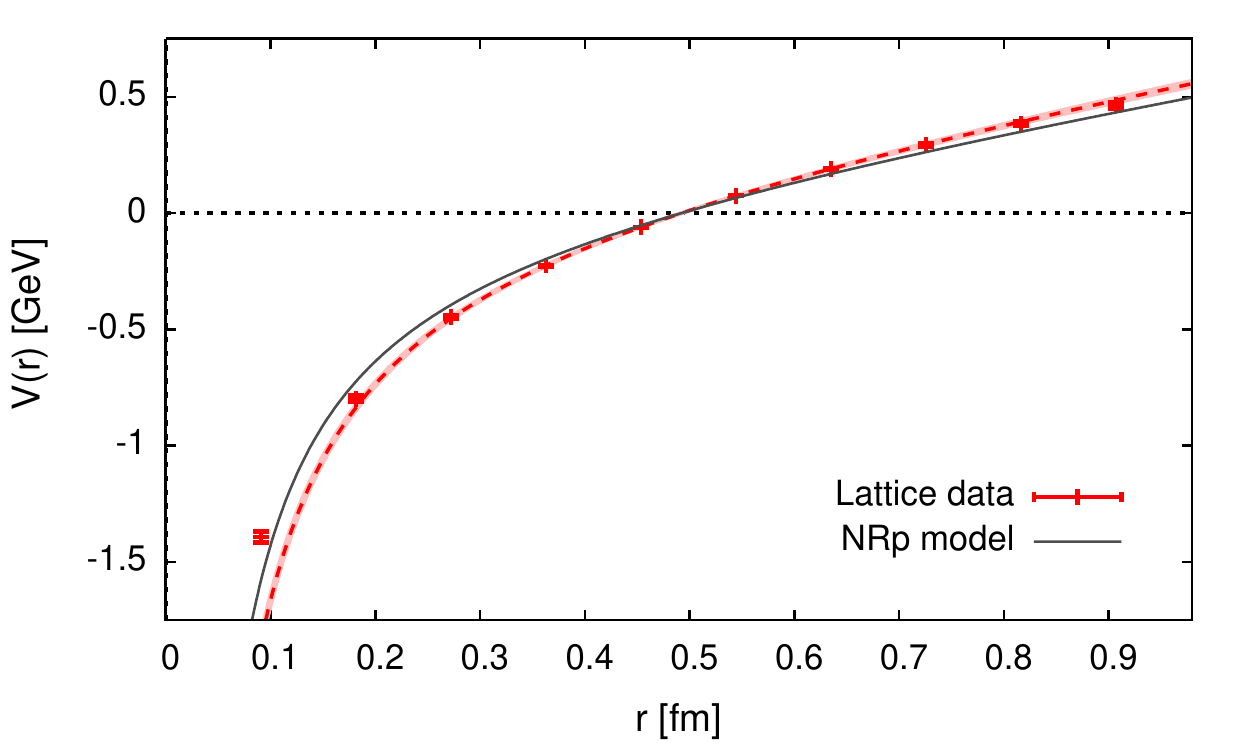}
       \includegraphics[width=.48\textwidth]{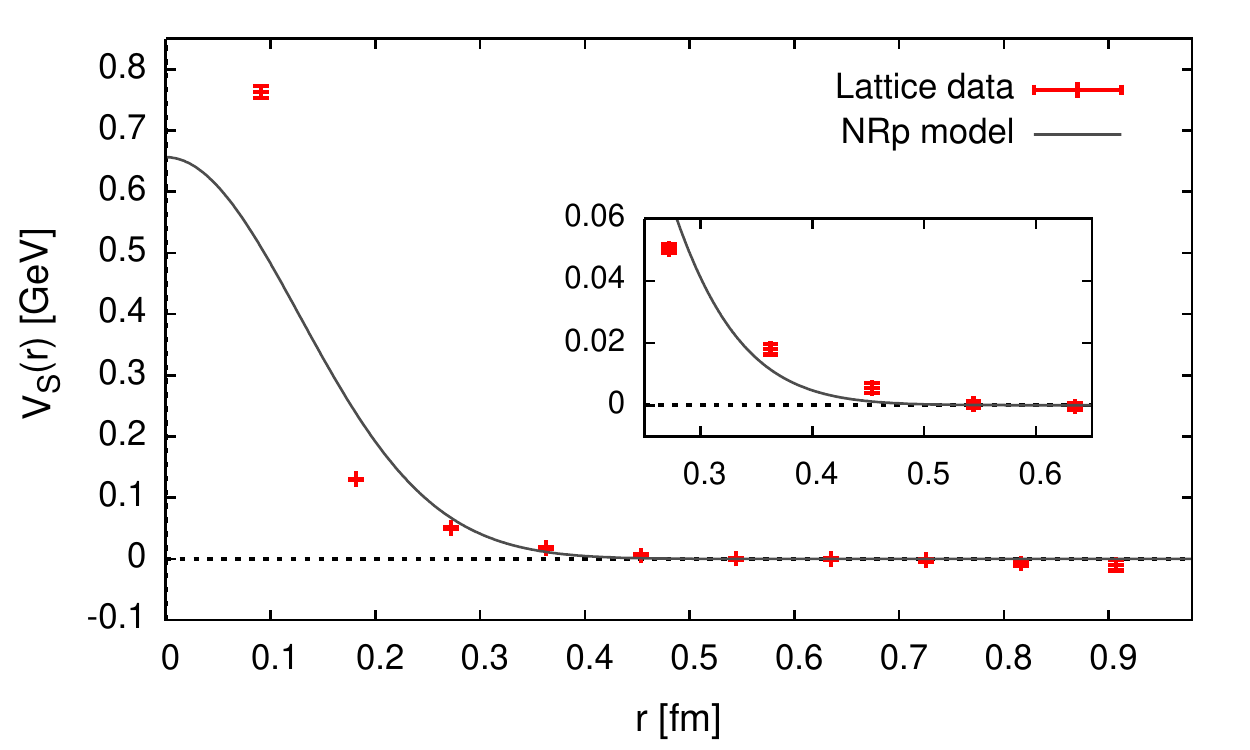}
   \caption{
      Spin-independent (left) and spin-dependent (right) 
      parts of the central charmonium potential.
      The solid curves show the phenomenological
      potentials adopted in a NRp model~\cite{Barnes:2005pb}.
      The dashed curve appeared in the left panel represent 
      a fit result with the Cornell parameterization.
      %The dashed curve and shaded band in the left panel represent 
      %a fit result with the Cornell parameterization and its statistical uncertainty 
      %calculated by the jack-knife method, respectively.
      \label{Pot_full}}
  \end{figure}
 First, we show a result of the spin-independent charmonium potential $V(r)$
 with the fitted curve (dashed curve) in Fig.~\ref{Pot_full}, where the constant term is subtracted
 to set $V(r_0)=0$ with the Sommer scale, $r_0\approx 0.5$~fm.
 The Cornell parametrization is also simply adopted for fit.
 We have carried out correlated $\chi^2$ fits
  with full covariance matrix for on-axis data over range
  $4\leq r/a \leq 10$.
  The fitting results are listed in Table~\ref{Tab_para}.
  The quoted errors represent only the statistical errors
  given by the jack-knife analysis.
  The phenomenological potential used in NRp models~\cite{Barnes:2005pb}
  is also plotted as a solid curve for comparison in Fig.~\ref{Pot_full}. 
  Although the charmonium potential obtained from
  lattice QCD is quite similar to the one in the NRp models,
  the string tension of the charmonium potential is slightly stronger
  than the phenomenological one.  Therefore our result indicates that
  there are only minor modifications required for the spin-independent
  central potential in the NRp models.
  Moreover, it seems that a large gap for the Coulombic coefficients between
  the conventional static potential and
  the phenomenological potential is filled by
  our new approach, where all orders of $1/\mQ$ corrections are
  nonperturbatively accounted for.
 \begin{table}%[th]
 \centering
   \caption{Summary of the Cornell parameters and the quark mass
     determined from lattice QCD. For comparison, the corresponding values
     adopted in a non-relativistic potential (NRp) model~\cite{Barnes:2005pb}
     are also included.
     \label{Tab_para}
   }
     \begin{tabular}{ccccc} \hline  \hline                                                        
       &This work& Polyakov lines & NRp model \\
       \hline
       $A$           & 0.861(17)    & 0.403(24) &0.7281 \\
       $\sqrt{\sigma}$ [GeV] & 0.394(7)    & 0.462(4) &0.3775 \\
       $\mQ$ [GeV]          & 1.74(3)    &  $\infty$       &1.4794 \\
       \hline  \hline                                                        
     \end{tabular}
 \end{table}

  In Fig.~\ref{Pot_full}, we next show the spin-spin term of
  the charmonium potential and the corresponding phenomenological one
  found in Ref.~\cite{Barnes:2005pb}. Our spin-spin potential exhibits
  the short range {\it repulsive interaction},
  which is required by the charmonium spectroscopy. 
  %, where the higher spin state in hyperfine multiplets receives heavier mass.
  It should be reminded that the Wilson loop approach
  fails to reproduce the correct behavior of the spin-spin interaction,
  since the leading-order spin-spin potential classified in pNRQCD
  becomes attractive at short distances~\cite{Koma:2006fw}.
  In contrast of the case of the spin-independent potential,
  the spin-spin potential obtained here is absolutely different
  from a repulsive $\delta$-function potential generated
  by perturbative one-gluon exchange. 
  Indeed, the finite-range spin-spin potential described
  by the Gaussian form is adopted in Ref.~\cite{Barnes:2005pb}, where
  many properties of conventional charmonium states at higher masses
  are predicted. 
   This phenomenological spin-spin potential is also plotted
  in Fig.~\ref{Pot_full} for comparison. There still remains a slight
  difference between the spin-spin potential from first principles QCD
  and the phenomenological one. In this sense, the reliable spin-dependent
  potential derived from lattice QCD can provide new and valuable
  information to the NRp models.
  
  \section{Summary}
  We have proposed the new method to determine
  the interquark potential at finite quark mass from lattice QCD.
  Using quenched lattice QCD, we have demonstrated
  that the spin-independent central potential defined in this method
  smoothly approaches the static $\QQbar$ potential given by Wilson loops
  in the infinitely heavy-quark limit.
  In dynamical lattice QCD simulations,
  we have studied both spin-independent and -dependent parts
  of the charmonium potential.
   The spin-independent charmonium potential obtained from lattice QCD
  with almost physical quark masses is quite similar to  the one used
  in the NRp models. The spin-spin potential properly
  exhibits the short range repulsive interaction. Its $r$-dependence,
  however, is slightly different from the
  phenomenological one adopted in Ref.~\cite{Barnes:2005pb}.
  Therefore, our charmonium potential derived from first principles QCD
   suggest that properties of higher-mass charmonium states predicted in the NRp models may change.
  
%   \section*{Acknowledgement}
    We acknowledge the PACS-CS collaboration and ILDG/JLDG for %~\cite{ILDG} for
  providing us with the gauge configurations. We would also like to
  thank H. Iida, Y. Ikeda and T. Hatsuda for fruitful discussions.
  This work was partially supported by JSPS/MEXT Grants-in-Aid
  (No.~22-7653, No.~19540265, No.~21105504 and No.~23540284).

\end{document}